\documentclass[twocolumn,floatfix,showpacs,superscriptaddress]{revtex4}
\usepackage{graphicx}
\usepackage{amsmath}
\usepackage{bm}
\begin{document}

\title{Quantum Goos-H\"{a}nchen effect in graphene}

\author{C. W. J. Beenakker}
\affiliation{Instituut-Lorentz, Universiteit Leiden, P.O. Box 9506, 2300 RA Leiden, The Netherlands}
\author{R. A. Sepkhanov}
\affiliation{Instituut-Lorentz, Universiteit Leiden, P.O. Box 9506, 2300 RA Leiden, The Netherlands}
\author{A. R. Akhmerov}
\affiliation{Instituut-Lorentz, Universiteit Leiden, P.O. Box 9506, 2300 RA Leiden, The Netherlands}
\author{J. Tworzyd{\l}o}
\affiliation{Institute of Theoretical Physics, Warsaw University, Ho\.{z}a 69, 00--681 Warsaw, Poland}

\date{December 2008}
\begin{abstract}
The Goos-H\"{a}nchen (GH) effect is an interference effect on total internal reflection at an interface, resulting in a shift $\sigma$ of the reflected beam along the interface. We show that the GH effect at a \textit{p-n} interface in graphene depends on the pseudospin (sublattice) degree of freedom of the massless Dirac fermions, and find a sign change of $\sigma$ at angle of incidence $\alpha^{\ast}=\arcsin\sqrt{\sin\alpha_{c}}$ determined by the critical angle $\alpha_{c}$ for total reflection. In an \textit{n}-doped channel with \textit{p}-doped boundaries the GH effect doubles the degeneracy of the lowest propagating mode, introducing a two-fold degeneracy on top of the usual spin and valley degeneracies. This can be observed as a stepwise increase by $8e^{2}/h$ of the conductance with increasing channel width.
\end{abstract}
\pacs{73.23.Ad, 42.25.Gy, 72.90.+y, 73.50.-h}
\maketitle

Analogies between optics and electronics have inspired the research on graphene since the discovery of this material a few years ago \cite{Nov04}. Some of the more unusual analogies are drawn from the field of optical metamaterials. In particular, negative refraction in a photonic crystal \cite{Not00} has an analogue in a bipolar junction in graphene if the width $d$ of the \textit{p-n} interface is less than the electron wave length $\lambda_{F}$ \cite{Che07}. Negative refraction is only possible for angles of incidence $\alpha$ less than a critical angle $\alpha_{c}$. For $\alpha>\alpha_{c}$ the refracted wave becomes evanescent and the incident wave is totally reflected with a shift $\sigma$ of order $\lambda_{F}$ along the interface. This wave effect is known as the Goos-H\"{a}nchen effect \cite{Goo47}, after the scientists who first measured it in 1947. The GH effect was already predicted in Newton's time and has become a versatile probe of surface properties in optics, acoustics, and atomic physics \cite{For01}. In particular, the interplay of the GH effect and negative refraction plays an important role in photonic crystals and other metamaterials \cite{Mar07,Tsa07}.

The electronic analogue of the GH effect has been considered previously \cite{Mil72,Fra74,Sin05,Che08}, including relativistic corrections, but not in the ultrarelativistic limit of massless electrons relevant for graphene. As we will show here, the shift of a beam upon reflection at a \textit{p-n} interface in graphene is strongly dependent on the sublattice (or ``pseudospin'') degree of freedom --- both in magnitude and sign. We calculate the average shift $\sigma$ after multiple reflections at opposite \textit{p-n} interfaces and (contrary to a recent expectation \cite{Zha08}) we find that $\sigma$ changes sign at $\alpha^{\ast}=\arcsin\sqrt{\sin\alpha_{c}}$. 

\begin{figure}[tb]
\centerline{\includegraphics[width=0.8\linewidth]{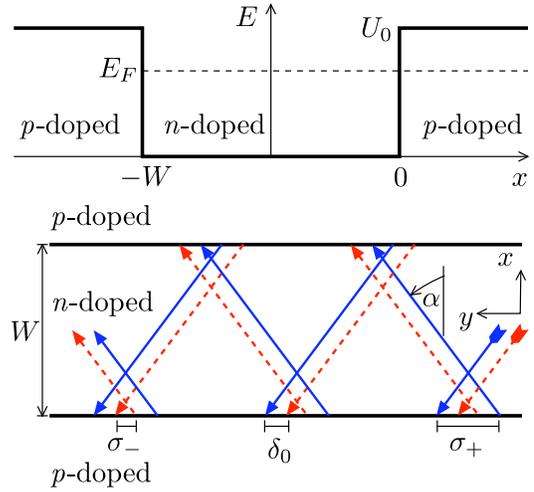}}
\caption{\label{fig_pnp}
Upper panel: Potential profile of an \textit{n}-doped channel between \textit{p}-doped regions. Lower panel: Top view of the channel in the graphene sheet. The blue solid line follows the center of a beam on the A sublattice, while the red dashed line follows the center on the B sublattice. The two centers have a relative displacement $\delta_{0}$. Upon reflection, each pseudospin component experiences alternatingly large and small shifts $\sigma_{\pm}$.
}
\end{figure}

In search for an observable consequence of the GH effect we study the conductance of the \textit{p-n-p} junction, for current \textit{parallel} to the interfaces (see Fig.\ \ref{fig_pnp}). (The conductance for current \textit{perpendicular} to the interfaces was calculated by Pereira et al.\ \cite{Per06}.) We find that the lowest mode in the \textit{n}-doped channel has a twofold degeneracy, observable as an $8e^{2}/h$ stepwise increase in the conductance as a function of channel width.

We recall some basic facts about the carbon monolayer called graphene \cite{Bee08,Cas09}. Near the corners of the Brillouin zone the electron energy depends linearly on the momentum, like the energy-momentum relation of a photon (but with a velocity $v$ that is 300 times smaller). The corresponding wave equation is formally equivalent to the Dirac equation for massless spin-$1/2$ particles in two dimensions. The spin degree of freedom is not the real electron spin (which is decoupled from the dynamics), but a pseudospin variable that labels the two carbon atoms (A and B) in the unit cell of a honeycomb lattice.

To calculate the GH shift we consider a beam,
\begin{equation}
\bm{\Psi}^{\rm in}(x,y)=\int_{-\infty}^{\infty}dq\,f(q-\bar{q})e^{iqy+ik(q)x}
\begin{pmatrix}
e^{-i\alpha(q)/2}\\
e^{i\alpha(q)/2}
\end{pmatrix},\label{Psiin}
\end{equation}
incident on a \textit{p-n} interface at $x=0$ from an \textit{n}-doped region $x<0$. The spinor wave function $\bm{\Psi}=(\Psi_{+},\Psi_{-})$ has pseudospin component $\Psi_{+}$ and $\Psi_{-}$ on the A and B sublattices. We require that $\bm{\Psi}^{\rm in}$ is a solution of the Dirac equation,
\begin{equation}
\left(-i\hbar v\sigma_{x}\frac{\partial}{\partial x}-i\hbar v\sigma_{y}\frac{\partial}{\partial y}+U\right)\Psi=E\Psi,\label{Diracequation}
\end{equation}
with $U=0$ (zero potential in the \textit{n}-doped region) and $E=E_{F}$ (the Fermi energy). This requirement fixes the dependence of the longitudinal wave vector $k$ and the angle of incidence $\alpha$ on the transverse wave vector $q$,
\begin{equation}
k=\sqrt{(E_{F}/\hbar v)^{2}-q^{2}},\;\;\alpha=\arcsin(\hbar vq/E_{F}). \label{kalphaqrelation}
\end{equation}
For brevity, we will set $\hbar v\equiv 1$ in some intermediate equations (restoring units in the final answers).

The transverse wave vector profile $f(q-\bar{q})$ of the beam is peaked at some $\bar{q}\in(0,E_{F}/\hbar v)$, corresponding to an angle of incidence $\bar{\alpha}=\arcsin(\bar{q}/E_{F})\in(0,\pi/2)$. None of our results depend on the shape of the profile, but for definiteness we take a Gaussian,
\begin{equation}
f(q-\bar{q})=\exp[-\tfrac{1}{2}(q-\bar{q})^{2}/\Delta_q^{2}],
\end{equation}
of width $\Delta_q$.

For $\Delta_{q}$ small compared to the Fermi wave vector $k_{F}=E_{F}/\hbar v$ we may expand $k(q)$ and $\alpha(q)$ to first order around $\bar{q}$, substitute in Eq.\ \eqref{Psiin}, and evaluate the Gaussian integral to obtain the spatial profile of the incident beam. At the interface $x=0$ the two components $\Psi_{\pm}^{\rm in}\propto\exp[-\tfrac{1}{2}\Delta_{q}^{2}(y-\bar{y}_{\pm}^{\rm in})^{2}]$ of $\bm{\Psi}^{\rm in}(0,y)$ are Gaussians of the same width $\Delta_{y}=1/\Delta_{q}$, centered at two different mean $y$-coordinates
\begin{equation}
\bar{y}^{\rm in}_{\pm}=\pm\tfrac{1}{2}\alpha'(\bar{q})=\pm\tfrac{1}{2}(k_{F}\cos\bar{\alpha})^{-1}.\label{xinpm}
\end{equation}
(The prime in $\alpha'$ indicates the derivative with respect to $q$.) The separation
\begin{equation}
\delta_{0}=|\bar{y}^{\rm in}_{+}-\bar{y}^{\rm in}_{-}|=(k_{F}\cos\bar{\alpha})^{-1}
\end{equation}
of the two centers is of the order of the Fermi wave length $\lambda_{F}=2\pi/k_{F}$, which is small compared to the width $\Delta_{y}$ but of the same order of magnitude as the GH shift --- so it cannot be ignored.

Similar considerations are now applied to the reflected wave,
\begin{equation}
\bm{\Psi}^{\rm out}=\int_{-\infty}^{\infty}dq\,f(q-\bar{q})e^{iqy-ik(q)x}r(q)
\begin{pmatrix}
-ie^{i\alpha(q)/2}\\
ie^{-i\alpha(q)/2}
\end{pmatrix},\label{Psiout}
\end{equation}
obtained from the incident wave \eqref{Psiin} by the replacements $k\mapsto -k$, $\alpha\mapsto\pi-\alpha$ and multiplication with the reflection amplitude $r(q)=|r(q)|e^{i\phi(q)}$. The two components $\Psi_{\pm}^{\rm out}$ of $\bm{\Psi}^{\rm out}(0,y)$ at the interface are Gaussians centered at
\begin{equation}
\bar{y}^{\rm out}_{\pm}=-\phi'(\bar{q})\mp\tfrac{1}{2}\alpha'(\bar{q})=-\phi'(\bar{q})\mp\tfrac{1}{2}(k_{F}\cos\bar{\alpha})^{-1}.\label{xoutpm}
\end{equation}

Comparison with Eq.\ \eqref{xinpm} shows that the first component of the spinor is displaced along the interface by an amount $\sigma_{+}=y_{+}^{\rm out}-y_{+}^{\rm in}=-\phi'(\bar{q})-\delta_{0}$, while the second component is displaced by $\sigma_{-}=y_{-}^{\rm out}-y_{-}^{\rm in}=-\phi'(\bar{q})+\delta_{0}$. The average displacement,
\begin{equation}
\sigma=\tfrac{1}{2}(\sigma_{+}+\sigma_{-})=-\phi'(\bar{q})=-{\rm Im}\,\frac{d}{dq}\ln r,\label{sigmadef}
\end{equation}
is the GH shift. As illustrated in Fig.\ \ref{fig_pnp}, after $N$ reflections the two components of the spinor are displaced by the same amount $N\sigma$ if $N$ is even and by a different amount $N\sigma\mp\delta_{0}$ if $N$ is odd. For $N\gg 1$ the difference $2\delta_{0}$ between the two displacements becomes small compared to the average shift $N\sigma$.

The formula \eqref{sigmadef} for the GH shift is generally valid for reflection from any interface. To apply it to the step function \textit{p-n} interface we calculate the reflection amplitude by matching $\bm{\Psi}^{\rm in}+\bm{\Psi}^{\rm out}$ at $x=0$ to the evanescent wave
\begin{align}
&\bm{\Psi}^{\rm ev}=\int_{-\infty}^{\infty}dq\,C(q)e^{iqy-\kappa(q)x}
\begin{pmatrix}
i(U_{0}-E_{F})\\
\kappa(q)+q
\end{pmatrix},\label{Psiev}\\
&\kappa=\sqrt{q^{2}-(\hbar v)^{-2}(E_{F}-U_{0})^2}.\label{kappadef}
\end{align}
This is a solution of the Dirac equation \eqref{Diracequation} (with $U=U_{0}$ and $E=E_{F}$) that decays into the \textit{p}-doped region $x>0$ for $\hbar v|q|>|E_{F}-U_{0}|$.

Continuity of the wave function at $x=0$ allows us to eliminate the unknown function $C(q)$ and to obtain the reflection amplitude,
\begin{equation}
r=\frac{ie^{i\alpha}(E_{F}-U_{0})+\kappa+q}{E_{F}-U_{0}+ie^{i\alpha}(\kappa+q)}.\label{rresult}
\end{equation}
The modulus $|r|=1$ for angles of incidence
\begin{equation}
\alpha>\alpha_{c}\equiv\arcsin|U_{0}/E_{F}-1|
\end{equation}
such that there is total reflection \cite{note1}. Substitution into Eq.\ \eqref{sigmadef} then gives the GH shift,
\begin{align}
\sigma&=\frac{\sin^{2}\alpha+1-U_{0}/E_{F}}{\kappa \sin\alpha\cos\alpha}
\nonumber\\
&=\frac{\lambda_{F}}{\pi\sin 2\alpha}\frac{\sin^{2}\alpha-{\rm sign}\,(U_{0}-E_{F})\sin\alpha_{c}}{\sqrt{\sin^{2}\alpha-\sin^{2}\alpha_{c}}}
.\label{sigmaresult}
\end{align}

\begin{figure}[tb]
\centerline{\includegraphics[width=0.8\linewidth]{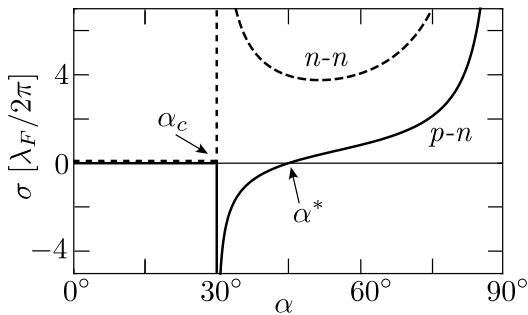}}
\caption{\label{fig_sigma}
Dependence on the angle of incidence $\alpha$ of the GH shift $\sigma$, calculated from Eq.\ \eqref{sigmaresult} for $U_{0}/E_{F}=1.5$ (solid curve, \textit{p-n} interface) and for $U_{0}/E_{F}=0.5$ (dashed curve, \textit{n-n} interface). The critical angle for total reflection (below which $\sigma=0$) equals $\alpha_{c}=30^{\circ}$ in both cases. The sign-change angle $\alpha^{\ast}=45^{\circ}$ for $U_{0}/E_{F}=1.5$.
}
\end{figure}

A negative GH shift (in the backward direction) appears at a \textit{p-n} interface (when $E_{F}<U_{0}$) for angles of incidence
\begin{equation}
\alpha_{c}<\alpha<\alpha^{\ast}\equiv\arcsin\sqrt{\sin\alpha_{c}}.
\end{equation}
For $\alpha>\alpha^{\ast}$ the GH shift is positive (in the forward direction), regardless of the relative magnitude of $E_{F}$ and $U_{0}$. In Fig.\ \ref{fig_sigma} we have plotted the $\alpha$ dependence of $\sigma$ for two representative cases.

The product $\sigma\cos\alpha\equiv\sigma_{\perp}$ is the shift in the direction perpendicular to the angle of incidence (while $\sigma$ is measured along the interface). This quantity becomes independent of $\alpha$ [in the interval $(\alpha_{c},\pi/2)$] when the charge density in the \textit{p}-doped region goes to zero at fixed charge density in the \textit{n}-doped region,
\begin{equation}
\sigma_{\perp}\rightarrow 1/k_{F}\;\;{\rm if}\;\;|E_{F}-U_{0}|\ll E_{F}\sin^{2}\alpha.\label{limit1}
\end{equation}
In this limit it does not matter for the sign of the shift if $E_{F}$ is larger or smaller than $U_{0}$. Since the perpendicular displacement of the two spinor components equals $\delta_{0}^{\perp}=\delta_{0}\cos\alpha=1/k_{F}$, the limit \eqref{limit1} for $\sigma_{\perp}$ implies that upon reflection one component has shift $\sigma_{\perp}-\delta_{0}^{\perp}=0$ equal to zero while the other component has shift $\sigma_{\perp}+\delta_{0}^{\perp}=2/k_{F}$.

As illustrated in Fig.\ \ref{fig_pnp}, the GH shift accumulates upon multiple reflections in the channel between two \textit{p-n} interfaces. If the separation $W$ of the two interfaces is large compared to the wave length $\lambda_{F}$, the motion between reflections may be treated semiclassically. The time between two subsequent reflections is $W/v\cos\alpha$, so the effect of the GH shift on the velocity $v_{\parallel}$ along the junction is given by
\begin{equation}
v_{\parallel}=v\sin\alpha + (\sigma/W)v\cos\alpha.
\end{equation}
Substitution of Eq.\ \eqref{sigmaresult} shows that, for $U_{0}>E_{F}$, the velocity $v_{\parallel}$ vanishes at an angle $\alpha^{\ast\ast}$ satisfying the equation
\begin{equation}
\sin^{2}\alpha^{\ast\ast}=(U_{0}/E_{F}-1)(\kappa W+1)^{-1},\label{alphaastast1}
\end{equation}
which for $k_{F}W\gg 1$ has the solution
\begin{equation}
\alpha^{\ast\ast}=\alpha_{c}+\frac{(1-\sin\alpha_{c})^{2}}{(k_{F}W)^{2}\sin 2\alpha_{c}\sin^{2}\alpha_{c}}+{\cal O}(k_{F}W)^{-4}.\label{alphaastast2}
\end{equation}

\begin{figure}[tb]
\centerline{\includegraphics[width=0.7\linewidth]{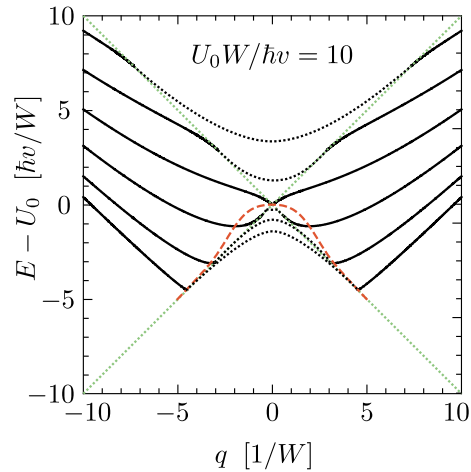}}
\caption{\label{fig_dispersion}
Energy $E$ of waves propagating with wave vector $q$ in the $y$-direction, bounded in the channel $-W<x<0$ by the potential profile in Fig.\ \ref{fig_pnp}. The different curves (black solid lines) correspond to different modes. (Only the six lowest channel modes are shown.) The curves are calculated from Eq.\ \eqref{kkappa} for $U_{0}W/\hbar v=10$ (semiclassical regime). The velocity $v_{\parallel}=dE/\hbar dq$ in the $y$-direction vanishes at the minima of the dispersion relation, given by Eq.\ \eqref{alphaastast1} (red dashed curve). At the (green) dotted lines $\hbar v|q|=|E-U_{0}|$ the channel modes are joined to modes in the wide region, as indicated schematically by the (black) dotted curves. 
}
\end{figure}

The vanishing velocity shows up as a minimum in the dispersion relation, obtained by solving the Dirac equation \eqref{Diracequation} with the potential profile
\begin{equation}
U(x)=\left\{\begin{array}{l}
U_{0}\;\;{\rm for}\;\; |x|>W/2,\\
0\;\;{\rm for}\;\; |x|<W/2.
\end{array}\right.
\label{U0def}
\end{equation}
Matching of propagating waves to decaying waves at $x=-W$ and $x=0$ produces the following relation between $E$ and $q$:
\begin{align}
&[q^{2}+E(U_{0}-E)]\sin kW+k\kappa\cos kW=0,\label{qErelation}\\
&k=\sqrt{E^{2}-q^{2}},\;\; \kappa=\sqrt{q^{2}-(U_{0}-E)^{2}}.\label{kkappa}
\end{align}

The dispersion relation $E(q)$ is plotted for the first few modes in Fig.\ \ref{fig_dispersion}. (A similar dispersion relation was obtained in Ref.\ \cite{Per06}.) The slope determines the velocity, $v_{\parallel}=dE/\hbar dq$. The minima in the dispersion relation where $v_{\parallel}=0$ are clearly visible for $E\lesssim U_{0}$. The locations of the minima are precisely \cite{note2} given by Eq.\ \eqref{alphaastast1} (red dashed curve). For $E\gtrsim U_{0}$ the GH effect increases the velocity, which is visible in the dispersion relation as a local increase in the slope of the dispersion relation. The solid curves in Fig.\ \ref{fig_dispersion} give the dispersion relation of modes that are confined to the narrow \textit{n}-doped channel. At the dotted lines $\hbar v|q|=|E-U_{0}|$ these channel modes are joined to the modes in the wide \textit{p}-doped region (as indicated by the dotted curves in Fig.\ \ref{fig_dispersion}). 

\begin{figure}[tb]
\centerline{\includegraphics[width=0.7\linewidth]{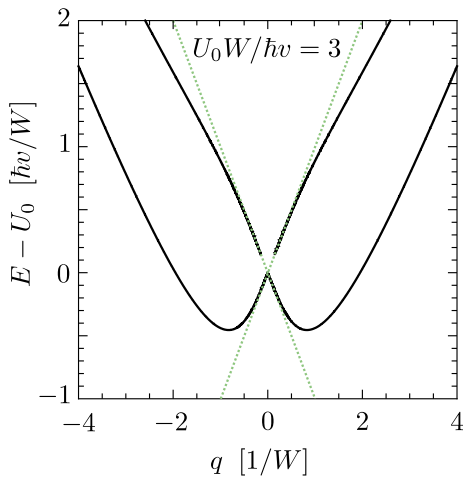}}
\caption{\label{fig_dispersion2}
Same as Fig.\ \ref{fig_dispersion}, but now showing the lowest channel modes in the fully quantum mechanical regime $U_{0}W/\hbar v=3$. The two minima at $q=\pm 0.83\,W^{-1}$ each contribute independently an amount of $4e^{2}/h$ to the conductance.
}
\end{figure}

\begin{figure}[tb]
\centerline{\includegraphics[width=0.8\linewidth]{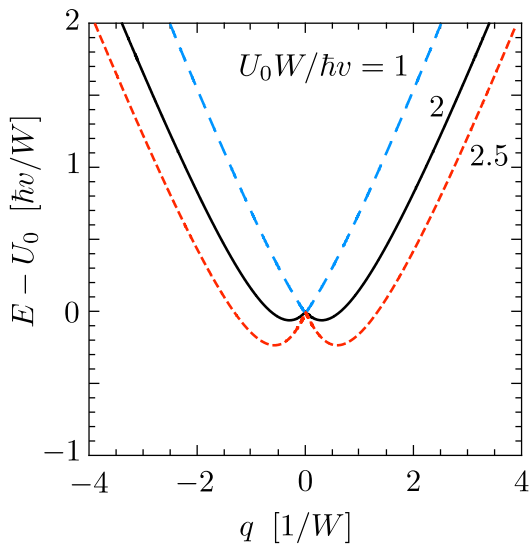}}
\caption{\label{fig_dispersion1}
Plot of the lowest mode for three values of $U_{0}W/\hbar v$, showing how the the two minima merge into a single minimum at $q=0$ upon reducing $W$.
}
\end{figure}

As the channel width is reduced so that $U_{0}W/\hbar v$ becomes of order unity, we enter the fully quantum mechanical regime. The minimum in the dispersion relation becomes very pronounced for the lowest channel mode, as we show in Fig.\ \ref{fig_dispersion2}. There are two minima at $q\approx 1/W$ and $q\approx -1/W$, each contributing to the conductance a quantum of $e^{2}/h$ per spin and valley degree of freedom. The total contribution to the conductance from the lowest channel mode is therefore $8e^{2}/h$. As shown in Fig.\ \ref{fig_dispersion1}, if $W$ is reduced further, the two degenerate minima in the dispersion relation merge into a single minimum at $q=0$ (this happens at $U_{0}W/\hbar v=1.57$), and for smaller $W$ the lowest channel mode again contributes the usual amount of $4e^{2}/h$ to the conductance.

\begin{figure}[tb]
\centerline{\includegraphics[width=0.9\linewidth]{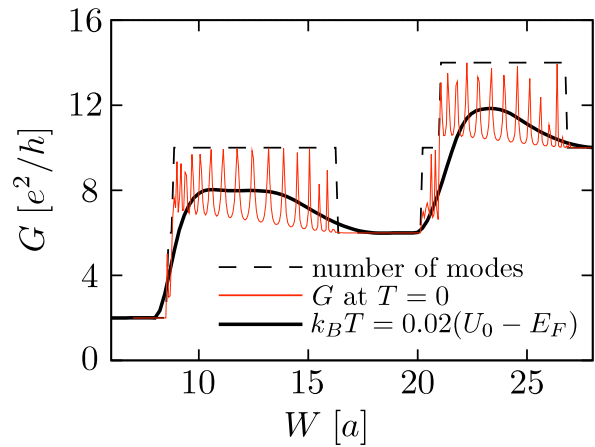}}
\caption{\label{fig_numerics}
Conductance versus channel width, calculated numerically at zero temperature (thin red curve) and at a finite temperature (thick black curve). The dashed black curve gives the number of propagating modes, calculated from the dispersion relation. 
}
\end{figure}

To test these analytical predictions, we have performed numerical simulations of electrical conduction in a tight-binding model of a graphene sheet covered by a split-gate electrode. The geometry is similar to that studied in Ref.\ \cite{Sny08} (but not in the \textit{p-n} junction regime of interest here). Using the recursive Green function technique on a honeycomb lattice of carbon atoms (lattice constant $a$) we obtain the transmission matrix $t$, and from there the conductance $G=(2e^{2}/h){\rm Tr}\,tt^{\dagger}$. Only the twofold spin degeneracy is included by hand as a prefactor, all other degeneracies follow from the simulation. The graphene strip is terminated in the $x$-direction by zigzag boundaries (separated by a distance $W_{\rm total}=220\,a$), while it is infinitely long in the $y$-direction. A smooth potential profile defines a long and narrow channel of length $L=1760\,a$ and a width $W$ which we vary between $0$ and $30\,a$. The potential rises from $0$ in the wide reservoirs (far from the narrow channel), to $U_{0}=0.577\,\hbar v/a$ underneath the gate, and has an intermediate value of $U_{\rm channel}=0.277\,\hbar v/a$ inside the channel (where the gate is split). The Fermi energy is kept at $E_{F}=0.547\,\hbar v/a$, so that it lies in the valence band underneath the gate, while it lies in the conduction band inside reservoirs and channel.

Results of the simulations are shown in Fig.\ \ref{fig_numerics}. From the dispersion relation we read off the total number of propagating modes (dashed curve). The zigzag edges of the graphene strip support one spin-degenerate edge mode, so the conductance levels off at $2e^{2}/h$ as the channel is pinched off. Upon widening the channel, the new channel modes have the $8$-fold degeneracy predicted by our analytical theory. The valley degeneracy is not exact (notice the small intermediate step at $W=20\,a$), as expected for a finite lattice constant. The zero-temperature conductance (thin red curve) shows pronounced Fabry-Perot type oscillations, due to multiple reflections at the entrance and exit of the channel, with an envelope that follows closely the number of propagating modes.

At finite temperature (black curve) the oscillations are averaged out, but the excess conductance characteristic of the Goos-H\"{a}nchen effect remains clearly observable at the temperature $T=0.02(U_{0}-E_{F})/k_{B}$ used in the simulation. Scaling up to realistic parameter values, we can set the channel width $W= 100\,{\rm nm}$ at the first conductance step, hence $U_{0}-E_{F}=0.03\,\hbar v/W\simeq 10\,{\rm K}$, so this would correspond to a temperature of $0.2\,{\rm K}$. The Fermi wave length $\lambda_{F}$ in the channel is of order $100\,{\rm nm}$ for these parameter values (of the same order as $W$ at the first step), well above the typical width $d\simeq 40\,\rm{nm}$ of a \textit{p-n} interface \cite{Hua07}. Note that $d$ is two orders of magnitude larger than $a=0.25\,{\rm nm}$, so the potential is indeed smooth on the scale of the lattice constant (as assumed both in the analytical and numerical calculations). For ballistic transport through the constriction the mean free path should be well above the $100\,{\rm nm}$ scale.

In conclusion, we have identified and analyzed a novel pseudospin-dependent scattering effect in graphene, that manifests itself as an $8e^{2}/h$ conductance step in a bipolar junction. This quantum Goos-H\"{a}nchen effect mimics the effects of a pseudospin degeneracy, by producing a pronounced double minimum in the dispersion relation of an \textit{n}-doped channel with \textit{p}-doped boundaries. Such a channel can be created electrostatically, and might therefore be a versatile building block in an electronic circuit.

This research was supported by the Dutch Science Foundation NWO/FOM.


\begin{thebibliography}{99}
\bibitem{note1} For $|\alpha|<\alpha_{c}$ one has $\kappa=-i\sqrt{(E_{F}-U_{0})^{2}-q^{2}}$ and $|r|<1$, because part of the incident wave is transmitted. One can readily check, by substitution into Eq.\ \eqref{sigmadef}, that $\sigma=0$ for these subcritical angles. As expected, there is no Goos-H\"{a}nchen shift without an evanescent wave.
\bibitem{note2} Although the equation \eqref{alphaastast1} for zero velocity was derived semiclassically, it is in fact quantum mechanically exact --- as we have found by differentiating the dispersion relation \eqref{qErelation} with respect to $q$ and searching for the value $q^{\ast\ast}=E\sin\alpha^{\ast\ast}$ at which the derivative vanishes.
\end{thebibliography}
\end{document}